\documentclass[aps,prd,twocolumn,superscriptaddress,showpacs,floatfix,preprintnumbers,nofootinbib]{revtex4-2}
\usepackage[utf8]{inputenc}

\usepackage[dvipsnames]{xcolor}      
\definecolor{lcolor}{rgb}{0.5,0,0}
\definecolor{citcolor}{rgb}{0,0,1}
\usepackage[breaklinks,colorlinks,urlcolor=blue,citecolor=citcolor,linkcolor=lcolor,linktoc=all]{hyperref}
\usepackage{color}
\usepackage{graphicx}	
\usepackage[utf8]{inputenc}
\usepackage{amsmath} 
\usepackage{amssymb}
\usepackage{slashed}

\newcommand{\mE}{m_{\text{E}}}
\newcommand{\mEl}{m_{\text{E,0}}}
\newcommand{\gE}{g_{\text{E}}}
\newcommand{\dA}{d_{\text{A}}}

\renewcommand{\tilde}{\widetilde}

\newcommand{\eq}{Eq.~}
\newcommand{\fig}{Fig.~}
\newcommand{\eqs}{Eqs.~}
\newcommand{\Sec}{Sec.~}
\newcommand{\app}{Appendix~}
\renewcommand{\Ref}{Ref.~}
\newcommand{\Refs}{Refs.~}
 
\newcommand{\sumint}{\hbox{$\sum$}\!\!\!\!\!\!\!\int} 
 
\newcommand{\sumintF}[1]{\sumint_{\{\widetilde{#1}\}}} 
 
\newcommand{\dLm}{\delta \mathcal{L}_m}
\newcommand{\Lbar}{\bar{\Lambda}}

\makeatletter
\g@addto@macro\bfseries{\boldmath}
\makeatother

\begin{document}
\title{Cool quark matter with perturbative quark masses}
\author{Tyler Gorda}
\email{tyler.gorda@physik.tu-darmstadt.de}
\affiliation{Technische Universit\"{a}t Darmstadt, Department of Physics, 64289 Darmstadt, Germany}
\affiliation{Helmholtz Research Academy for FAIR, 64289 Darmstadt, Germany}
\author{Saga S\"appi}
\email{ssappi@ectstar.eu}
\affiliation{European Centre for Theoretical Studies in Nuclear Physics and Related Areas (ECT*) and Fondazione Bruno Kessler, Strada delle Tabarelle 286, I-38123, Villazzano (TN), Italy}
\date{\today}

\begin{abstract}
In perturbative thermal calculations, introducing nonzero quark masses causes considerable complications. In this article we describe an approximation scheme valid for sufficiently light masses which significantly simplifies the relevant calculations with only a small loss in accuracy. We apply the scheme to Quantum Chromodynamics, with two massless and one massive flavor, obtaining analytic results which are in excellent agreement with numerical non-approximated results.  Our results are accurate to $O(g^5)$ in the strong coupling parameter $g$, as long as either the chemical potential $\mu$ of the quark species with nonzero mass $m$ satisfies $m = O(g \mu)$ or if the temperature $T$ satisfies $m = O(g T)$. We find that these conditions are always satisfied for the three lightest quarks for the values of $\mu$, $T$ where Quantum Chromodynamics is perturbatively convergent.
\end{abstract}

\maketitle

\section{Introduction}
The pressure (or equation of state) of Quantum Chromodynamics (QCD) is the most fundamental equilibrium quantity in the study of strongly interacting matter. Knowledge of the pressure is of fundamental importance in studies of the early universe \cite{Schwarz:2003du,fodor}, heavy-ion collisions \cite{Busza:2018rrf}, and neutron stars (NSs) \cite{Lattimer:2000nx}. At sufficiently large temperatures $T$ or chemical potentials $\mu$, asymptotic freedom allows one to expand the pressure in a weak-coupling series in the strong coupling $g$. Perturbative QCD (pQCD) calculations making use of this property have been applied to compute the pressure to partial next-to-next-to-next-to-leading order (N3LO) at $T>0,\mu=0$ \cite{Kajantie:2002wa} and at $T=0,\mu>0$ \cite{Gorda:2021znl,Gorda:2021kme}, and to complete next-to-next-to-leading order (N2LO) at both $T>0,\mu>0$ \cite{Kurkela:2016was}. 

However, in many such calculations, the masses of quarks $m$ are neglected as small. While there is evidence suggesting that the effect of quark masses to the pressure are indeed small \cite{Kurkela:2009gj}, there are \emph{also} indications that they could play a role in certain systems, such as hypothetical strange quark stars \cite{Fraga:2004gz}. 
In fact, even in NSs the quark-mass effects have particularly interesting qualitative implications:  Without masses, QCD matter with equal densities of up, down, and strange quarks is both $\beta$-equilibriated and charge neutral, which are precisely the conditions experienced for cold, isolated neutron stars. Thus, in order to include leptons in $\beta$-equilibrium at high densities, the strange-quark-mass effects are necessary. 

Including quark masses in full generality within pQCD calculations is a tedious process. It has been achieved at $T=0,\mu>0$ to N2LO \cite{Kurkela:2009gj} and at $T>0,\mu>0$ to next-to-leading order (NLO) \cite{Laine:2006cp}, but even in these relatively low-order cases the resulting expressions are cumbersome and involve complicated numerical integrals: Indeed, even in the simplest leading order (LO) and NLO cases the massive, nonzero-temperature one-loop integrals cannot be expressed in terms of standard special functions and their derivatives.

Here, we choose a different approach, and examine a method of obtaining vastly simpler expressions by sacrificing only very little accuracy. We start with the observation that at the values of $T$ and $\mu$ where pQCD results are reliable, the ratio $m/(\pi T)$ or $m/\mu$ is still small for the three lightest quarks.\footnote{Note that with regards to temperature, the relevant scale is the energy of the first fermionic  Matsubara mode, which is $\pi T$.} For the more massive quarks, this approximation breaks down, and in particular one may not use this approximation when studying quark decoupling \cite{Laine:2006cp}. However, in many situations of interest, only the three lightest quarks are active.
For instance, at high densities the pQCD results are convergent for $\mu_q \gtrsim 0.9$~GeV, which is larger than the mass of the strange quark $m_s\approx 0.1$~GeV, and the charm quark is not yet active at this density \cite{Jimenez:2019kji}. A similar situation exists for temperatures $T \gtrsim 0.3$~GeV, where $m_s/(\pi T)$ is small, which approaches the regime of relevance for heavy-ion collisions. 

In this paper, we compute the pQCD pressure at nonvanishing chemical potentials, temperatures, and quark masses to $O(g^5)$ by expanding it in the quark masses, under the simplifying assumption that the masses are \emph{soft} --- here defined as $O(g\mu)$ or $O(g T)$. While there is no fundamental mechanism to connect the quark mass and coupling in this way, we observe that the mass of the strange quark is indeed a soft quantity\footnote{Strictly speaking, we demand the ratio $m_q/m_\text{E}$ to be at most order unity, where $\mE$ [defined below in \eq\eqref{eq:leading_me}] is the scale of Debye screening of gluons, but we opt to use the conventional notation for soft scales instead.} for the values of $\mu$ and $T$ discussed in the previous paragraph. Moreover, by considering a more general scaling of the quark masses in powers of the coupling, we can organize the relative importance of the mass corrections and coupling corrections within the same series expansion. Our result then brings the massive pQCD result at both $T, \mu > 0$ in line with the $m=0$ counterpart. 

\begin{figure*}[hbt]
    \centering
    \includegraphics[width=0.96\textwidth]{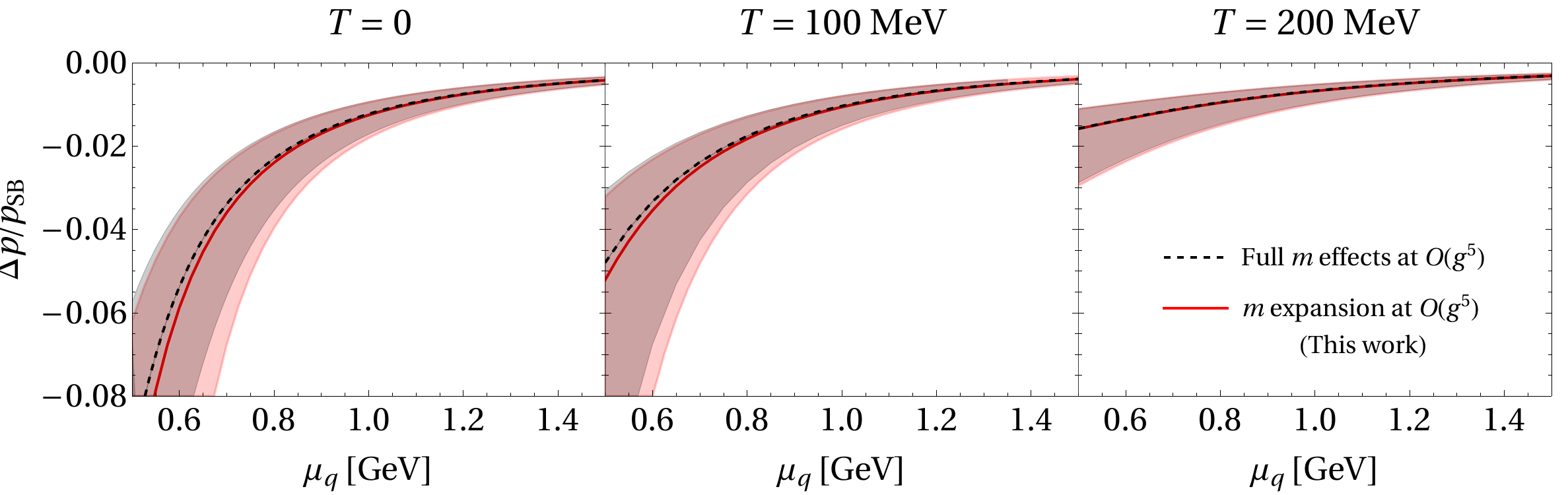}
    \caption{A comparison of the mass correction to the pressure $\Delta p \equiv - \Delta \Omega$ calculated with the full quark-mass dependence (dashed lines) and within our perturbative scheme (solid lines) at $T = 0$ (left panel), $T = 100$~MeV (middle panel), and $T = 200$~MeV (right panel)  for three quarks with the same chemical potential $\mu_q$. We take the up and down quarks as massless, and the one-loop running strange quark mass is used. 
    In the leftmost panel, the comparison to the full results from \Ref\cite{Fraga:2004gz} is used, while in all other panels, the full result from \Ref\cite{Laine:2006cp} plus the full mass correction to the DR term, computed in the appendices is shown. 
    All panels show the mass correction divided by the free massless pressure, and the shaded regions show the renormalization-scale variation, as explained in the main text.}
    \label{fig:pres_comparisons_prog}
\end{figure*}

\begin{figure}[hbt]
    \centering
    \includegraphics[width=0.4\textwidth]{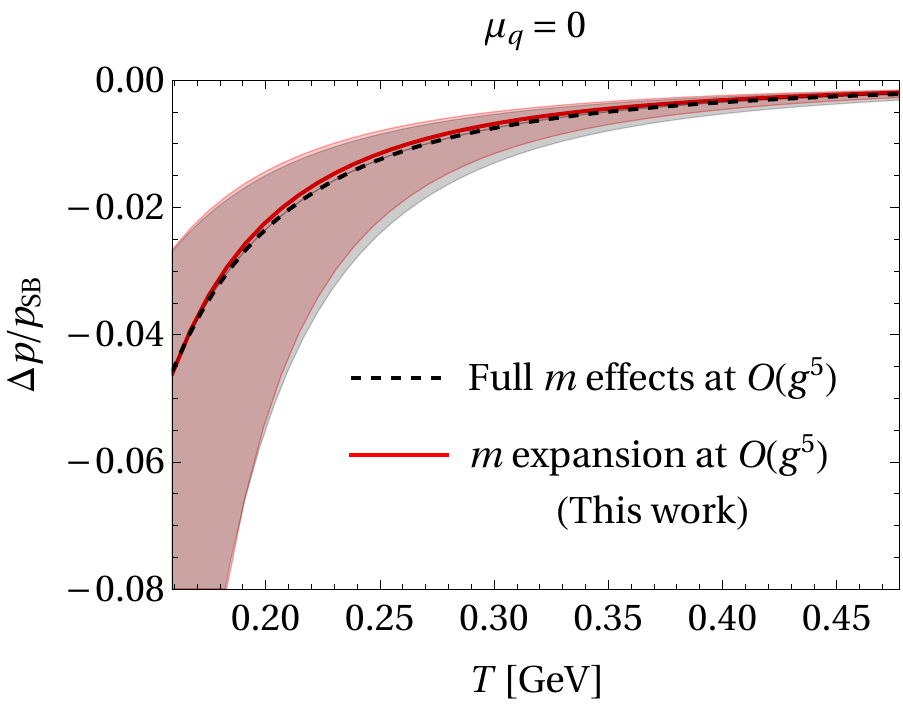}
    \caption{A comparison of the mass corrections to the pressure $\Delta p \equiv - \Delta \Omega$ calculated with the full quark-mass dependence (dashed lines) and within our perturbative scheme (solid lines) as a function of $T$ at $\mu_q = 0$. See \fig\ref{fig:pres_comparisons_prog} for more details.}
    \label{fig:pres_comparisons_mu0}
\end{figure}

The outline of our paper is as follows. 
In \Sec\ref{sec:methodology}, we start by explaining how the assumption of soft masses allows us to treat the quark-mass term in the Dirac Lagrangian as a perturbation and expand expressions about the massless limit, as well as outline the structure of the free energy after an expansion in soft quark masses. Expanding in the masses allows us to consider significantly simpler loop integrals and obtain closed-form results in $3+1$ dimensions, which we explicitly demonstrate in \Sec\ref{sec:calc}, giving additional details in the appendices.
We next in \Sec\ref{sec:results} examine the effect of these mass corrections to the pressure by comparing our simple results with the exact numerical expressions, finding good agreement. We then provide some applications of our results to the astrophysically relevant regime of high density and small to moderate temperatures, evaluating quantities which all become trivial in the limit of vanishing quark masses. Finally, we conclude with a small discussion.

\begin{figure*}
    \centering
    \includegraphics[width=0.98\textwidth]{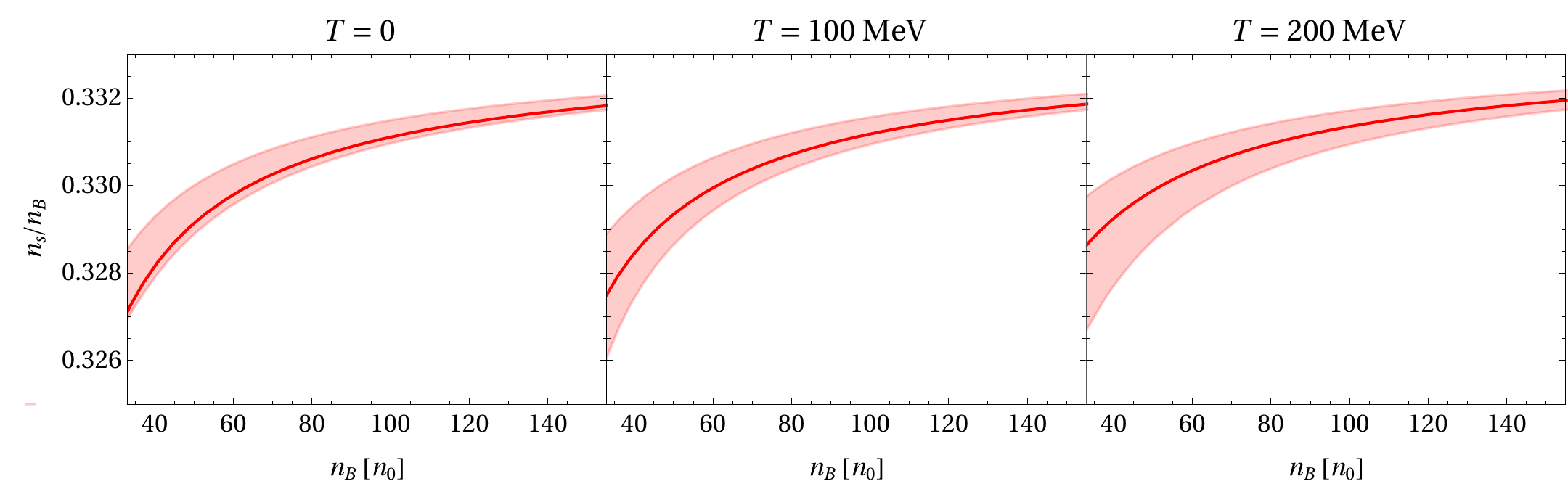}
    \caption{Left: A comparison of the strangeness fraction $n_s / n_{\rm B}$ at $T = 0$ (left panel), $T = 100$~MeV (middle panel), and $T = 200$~MeV (right panel)  for charge-neutral matter in beta equilibrium, ignoring neutrinos. Here, $n_0 = 0.16\text{ fm}^{-3}$ corresponds to nuclear saturation density. Note that ignoring the strange quark mass would set $n_s / n_{\rm B} = 1/3$ under these conditions. }
    \label{fig:xs_comparisons}
\end{figure*}

\section{Methodology} 
\label{sec:methodology}
Conceptually, our method to compute the (vacuum-subtracted) QCD grand potential $\Omega_\text{QCD}$ of matter with nonzero $\mu$, $T$, and $m$ is as follows. For simplicity in this section, let us assume only one quark flavor with mass $m$. We consider the massive QCD Lagrangian, and separate off the quark-mass part
\begin{equation}
    \mathcal{L}_{\text{QCD,} m} = \mathcal{L}_{\text{QCD,} m = 0} + m \bar{\psi} \psi.
\end{equation}
We will consider the quark-mass term in the above as a perturbation $\delta \mathcal{L}_m \equiv m \bar{\psi} \psi$ and calculate corrections to the grand potential as insertions of this mass term.
Such a calculational scheme is justified for massive particles for all but the $n = 0$ Matsubara mode of a massive boson; in particular, it is always justified for fermions as long as $m/\mu$ or $m/(\pi T)$ is small. This condition is always fulfilled in the case of the three lightest quark flavors for the region of the $T,\mu$-plane where pQCD exhibits good convergence. Thus, it is only in the vacuum subtractions where we must compute expressions unexpanded in the mass, and everywhere else we can use simpler, expanded expressions.\footnote{These vacuum subtractions will all contribute at $O(g^n m^4)$ on dimensional grounds (with the powers of the coupling present for loop corrections).} In fact, only even powers of $\dLm$ will lead to nonzero contributions, since one insertion of $\dLm$ adds one more fermion propagator to a fermion loop without another $\gamma^\mu$ from a new vertex, leading to a zero trace.

Let us now determine precisely what terms contribute to the grand potential of QCD matter under the conditions described above. The grand potential has contributions from three dynamical scales: the hard [$O(T,\mu)$], soft [$O(g \mu, g T)$] and ultrasoft [$O(g^2 T)$] scales. Contributions from the hard scale can be evaluated using unresummed perturbation theory, while the soft and ultrasoft contributions for $T>0$ can be treated using the dimensionally reduced (denoted ``DR'')  effective field theories of electrostatic QCD (EQCD) and magnetostatic QCD (MQCD), the latter of which does not enter until $O(g^6)$ and as such will not be needed explicitly here. Lastly, the soft contributions from $\mu > 0$ are associated with hard-thermal loop (HTL) perturbation theory. 

Gathering together the contributions from different sources, we have the following expression, wherein we must subtract out the unresummed (``naive'') contributions from each of these theories to avoid potential double counting \cite{Kurkela:2016was}
\begin{equation}
    \Omega_{\rm QCD} = \Omega^{\text{naive}}_\text{QCD} + \underbrace{\left( \Omega_\text{DR}^\text{res} - \Omega_\text{DR}^\text{naive} \right)}_{\Omega_\text{DR}^\text{corr}} + \underbrace{\left ( \Omega_\text{HTL}^\text{res} - \Omega_\text{HTL}^\text{naive} \right )}_{\Omega_\text{HTL}^\text{corr}},
\end{equation}
where here $\Omega^{\text{naive}}_\text{QCD}$ is our notation for the unresummed full-theory diagrams. These contributions have the following expansions in $g$ up to $O(g^6)$ \cite{Vuorinen:2003fs,Kurkela:2016was}\footnote{We have been notified by the authors of \cite{Kurkela:2016was} that the zero-temperature limit given in the publication has a minor error, and the first term of the second line of their \eq(B7) should have an overall factor of $1/48$ instead of $1/72$. This only applies to their zero-temperature results, which we have not directly used.} 
\begin{align}
    \label{eq:Omega_naive}
    \Omega_\text{QCD}^\text{naive} =& \; \Omega_\text{free} + g^2 \Omega_{2,\text{QCD}} + g^4 \Omega_{3,\text{QCD}} + O(g^6) \displaybreak[0]\\
    \label{eq:Omega_DR}
    \Omega_\text{DR}^\text{corr}/T =& \; \mE^3 \Omega_{1,\text{DR}} + \gE^2 \mE^2 \Omega_{2,\text{DR}} + \gE^4 \mE \Omega_{3,\text{DR}} \nonumber \\ &+ O(g^6) \displaybreak[0]\\
    \Omega_\text{HTL}^\text{corr} =& \; \mE^4 \Omega_{1,\text{HTL}} + O(g^6).
\end{align}
For a general description of the dimensionally reduced sector, see \Ref\cite{Kajantie:2002wa}. Above, the perturbative expansions of the EQCD parameters are given (in $d$ = 3) by
\begin{align}
    \mE^2 =& g^2 \alpha_\text{E4} + \frac{g^4}{(4\pi)^2} \alpha_\text{E6} + O(g^6) \\
    \gE^2 =& g^2 T + O(g^4),
\end{align}
where the $\alpha_{\mathrm{E}i}$ are matching parameters of EQCD, often written at $\mu=0$ but obtainable for $\mu>0$ via their general definitions \cite{Laine:2005ai}. For massive quarks, the $\Omega$ and $\alpha_\text{E}$ functions appearing above will in general become functions of this quark mass as well. Let us define 
\begin{equation}
    \Delta \Omega(m,\mu,T) = \Omega_{\rm QCD}(m,\mu,T) - \Omega_{\rm QCD}(0,\mu,T)
\end{equation} 
as the contribution from the quark masses, with the massless part in this expression being given in \Ref\cite{Kurkela:2016was}. Due to the structure of the loop integrals at nonzero $T,\mu$, the nonzero-mass expressions are expected to be analytic in the squared masses (beyond logarithms related to renormalization). We use this assumption schematically in our power-counting, and observe it \emph{a posteriori} at low orders.  Hence, after expanding for soft $m$, we have the following contributions to $\Omega$ up to $O(g^5)$. For the unresummed QCD pressure:
\begin{enumerate}
    \item $\Omega_\text{free}$ to $O(g^0 m^4)$,
    \item $g^2 \Omega_{2,\text{QCD}}$ to $O(g^2 m^2)$,
\end{enumerate}
and for the DR term:
\begin{enumerate}
    \item $\mE^3 \Omega_{1,\text{DR}}$ to $O(g^3 m^2)$.
\end{enumerate}
As the mass corrections from the HTL sector will not contribute until $O(g^4 m^2)$, they are not necessary here.

\section{Evaluating the massive QCD free energy to \texorpdfstring{$O(g^5)$}{O(g\^{}5)}}
\label{sec:calc}

We start by evaluating the unresummed contributions. These come in two types. First, there are medium-corrections to the one- and two-loop hard-theory diagrams, which can be calculated perturbatively in the quark mass $m$. Second, there are vacuum corrections from the quark mass, which cannot be expanded in the quark mass $m$. The latter is, by dimensional arguments, $O(m^4)$, so only the free contribution is needed here.

At this point, we may also immediately generalize the earlier discussion to $N_f$ quarks with different masses $ m_f $ and chemical potentials $ \mu_f$. Using the superscripts ``M'' and ``V'' to distinguish the matter and vacuum parts, we find the following corrections in the quark masses, leaving the details for \app\ref{app:unres_QCD}:  
\begin{align}
    \label{eq:Domega_MFD}
    \Delta\Omega_\text{free}^\text{M} \simeq& -2 N_c \sum_f \left ( 
    m_f^2 \widetilde{\mathcal{I}}^0_1 
    - \frac{m_f^4}{2} \widetilde{\mathcal{I}}^0_2 
    \right ) \displaybreak[0]\\
    \label{eq:Domega_VFD}
    \Delta\Omega_\text{free}^\text{V} =& 
    -2 N_c \sum_f \int_P \ln (P^2 + m_f^2) \nonumber \\
    =& - \frac{4 N_c}{D} \left( \frac {e^{\gamma_{\rm E}} \Lbar^2}{4 \pi} \right)^\epsilon \frac{\Gamma ( 1 - \frac{D}{2} )}{(4 \pi)^{D/2}} \sum_f (m_f^2)^{\frac{D}{2}} \nonumber \\
    \simeq& \sum_f \frac{N_c m_f^4}{(4 \pi)^2} \left[ \frac{1}{\epsilon} + \frac{3}{2} + 2 \ln \frac{\Lbar}{m_f} + O(\epsilon) \right], \displaybreak[0]\\
    \label{eq:Domega_2}
    g^2 \Delta\Omega_\text{2,QCD} =& 
    -2 \dA g^2 \sum_f m_f^2 \left[ 
        \tilde{\mathcal{T}} 
        + (1 - \epsilon ) (\widetilde{\mathcal{I}}^0_1 - \mathcal{I}^0_1) \widetilde{\mathcal{I}}^0_2 
    \right ],
\end{align}
where $\gamma_{\rm E} = 0.5772\ldots$ is the Euler--Mascheroni constant and $\Lbar$ is the $\overline{\text{MS}}$ renormalization scale. The sums are over all quark flavors, and we note that the integrals $\widetilde{\mathcal{I}}_n^k$ as well as the two-loop integral $\widetilde{\mathcal{T}}$, whose explicit expressions are given in \app\ref{app:ints}, each depend on a (single) chemical potential $\mu_f$ as well as the temperature $T$. The one remaining contribution at $O(g^5)$ is the first correction to $\mE$ from a nonzero quark mass, or equivalently the first mass correction to the matching coefficient $\alpha_\text{E4}$.

\begin{figure*}
    \centering
    \includegraphics[width=0.98\textwidth]{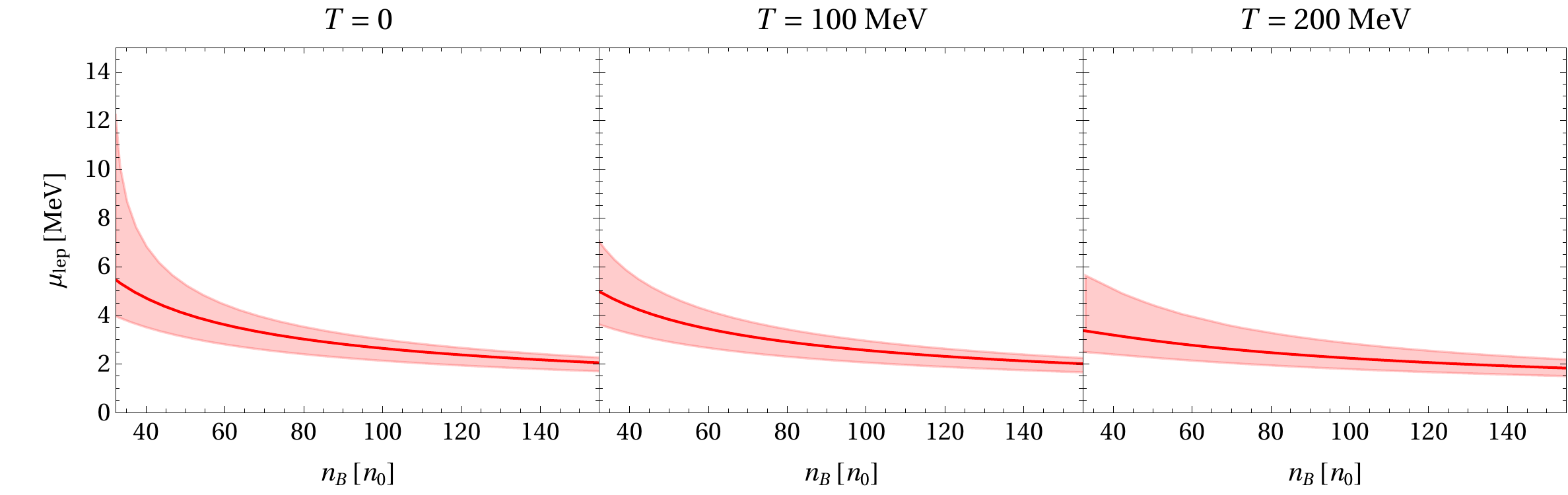}
    \caption{Left: A comparison of lepton chemical potential $\mu_\text{lep}$ at $T = 0$ (left panel), $T = 100$~MeV (middle panel), and $T = 200$~MeV (right panel)  for charge-neutral matter in beta equilibrium, ignoring neutrinos. Note that ignoring the strange quark mass would set $\mu_\text{lep} = 0$ under these conditions. }
    \label{fig:mu_lep}
\end{figure*}

We leave details of this computation to \app\ref{app:mass_exp_DR}, including writing down expressions that can be used to compute numerical results for any $m_f$. In the end the computation results in
\begin{equation}\label{eq:mEcorr}
    \Delta(\mE^3 \Omega_\text{1,DR}) 
    = \frac{2 \dA}{(4 \pi)^3} \mEl g^2 \sum_f m_f^2, 
\end{equation}
with the leading-order $\mEl^2$ given by \cite{Laine:2005ai}
\begin{equation}
    \mEl^2 = (d-1) g^2 \left \{ (d-1) N_c \mathcal{I}_1^0 + \sum_f \left[ 2\widetilde{\mathcal{I}}_2^2 -\widetilde{\mathcal{I}}_1^0 \right] \right \}.
\label{eq:leading_me}
\end{equation}
Note that we have written this definition of $\mEl$ in a form that is conveniently generalized to nonzero quark mass, instead of using zero-mass recurrence relations for the scalar integrals $\widetilde{\mathcal{I}}_n^k$, which are commonly applied in similar contexts.

\section{Results and Discussions}
\label{sec:results}

The sum of \eqs\eqref{eq:Domega_MFD}-\eqref{eq:mEcorr} constitute our unrenormalized result, with the masses $m_f$ appearing in these equations being the bare quark masses $m_{f,\text{B}}$. To carry out renormalization and cancel the divergences present in $\Delta \Omega_\text{free}^\text{V}$ in \eq\eqref{eq:Domega_VFD}, as well as in $\widetilde{\mathcal{T}}$ and $\widetilde{\mathcal{I}}_2^0$, which can be derived from \eqs\eqref{eq:Ttilde}-\eqref{eq:Itilde}, we use the standard one-loop relation 
\begin{equation}
    m_{f,\text{B}} = \left( 1 + \frac{\delta_1 g^2}{(4 \pi)^2} + \cdots \right) m_{f,\text{ren}},
\end{equation}
with $\delta_1 = -3 C_\text{F} / \epsilon$, in the above equations, which is only necessary in the terms proportional to $m_f^2$. This removes the remaining divergences, and we can then use our results with one-loop running quark masses $m_f(\Lbar)$ and two-loop running $\alpha_s(\Lbar)$. Two-loop running of the coupling is necessary since our results are accurate beyond $O(g^4)$, while one-loop running of the mass is sufficient since we are only working to $O(m^4)$. We henceforth work with the three lightest quarks only, taking $m_u = m_d = 0$, and using the one-loop running of the strange quark mass, with $m_s(2 \text{ GeV}) = 93.6$~MeV, which is the central value from \Ref\cite{Chakraborty:2014aca}, and we take $\Lambda_\text{QCD} = 378$~MeV in the two-loop running of the coupling. This quark content is relevant for the environment connecting to NS cores and the mergers of NS-NS binaries.

We first compare our renormalized results to the full results obtained without expanding in the quark mass. Since we are only calculating contributions up to $O(g^5)$ assuming soft $m$, we can take as the full result the first two terms of the unresummed $\Omega_\text{QCD}^\text{naive}$ in \eq\eqref{eq:Omega_naive} and the leading term in $\Omega_\text{DR}/T$ in \eq\eqref{eq:Omega_DR} both with the full quark-mass dependence, since including the quark mass anywhere else will only result in higher-order corrections. The unresummed contributions to this order were calculated for all $T$ in \Ref\cite{Laine:2006cp}, with the simpler $T = 0$ limit calculated in \Ref\cite{Fraga:2004gz}. The DR correction is calculated below in \app\ref{app:mass_exp_DR}. The comparison with our results is shown in \fig\ref{fig:pres_comparisons_prog}, for quark matter at high quark chemical potential and increasing temperatures, and in \fig\ref{fig:pres_comparisons_mu0} for a hot quark-gluon plasma at zero quark chemical potential. In these figures, we show the usual renormalization-scale variation in $\Lbar = X \sqrt{(2\mu_q)^2+ (0.723 \times 4 \pi T)^2}$ \cite{Kurkela:2016was,Kurkela:2009gj,Kajantie:2002wa}, with $X \in \{1/2, 1, 2\}$, with the central dashed or solid lines corresponding to $X = 1$.  As these figures illustrate, our analytic perturbative results are in excellent agreement with the full, numerical expressions at all temperatures and densities where pQCD is applicable.

We next apply our results to investigate three quantities of approximate astrophysical interest. When plotting these quantities, we show results down to the density where the calculated pQCD pressure diverges and becomes negative. 
In \fig\ref{fig:xs_comparisons}, we show the strangeness fraction $n_s/n_B$ as a function of baryon chemical potential ${\mu_B = \mu_u+\mu_d+\mu_s}$ for charge-neutral matter at high density in beta equilibrium at three different temperatures. Here, and below $n_i$ is the number density of the relevant particle $i$, and $n_B$ is the baryon density ${n_B = n_u + n_d + d_s}$. We take as the possible matter content the three lightest quarks $u, d, s$ and the two lightest leptons $e, \mu$, with the leptons being taken as non-interacting beyond the condition of beta equilibrium. We neglect the mass of the electron in the following, but we include the mass of the muon (and do not expand in it). We implement beta equilibrium and charge neutrality by the usual two conditions
\begin{align}
    \label{eq:beta_equil}
    \mu_d &= \mu_s = \mu_u + \mu_\text{lep}, \quad \mu_e = \mu_\mu = \mu_\text{lep} \\
    &\frac{2}{3} n_u- \frac{1}{3} n_d - \frac{1}{3} n_s - n_e - n_\mu = 0,
\end{align}
and so we explicitly ignore (i) the presence of neutrinos and neutrino trapping, and (ii) any possible modifications to the $T = 0$ equilibrium conditions that may arise from the blurring of the Fermi surfaces as the temperature is increased \cite{Alford:2018lhf}. The former condition can be easily weakened by including neutrino chemical potentials in \eq\eqref{eq:beta_equil}, while the latter requires a careful calculation of rates. In general, both of these conditions must be included to correctly model the full astrophysical environment of binary NS-NS mergers. We see from this figure that the strangeness fraction deviates from its high-density limit of $1/3$ by less than $2\%$ at all of the densities and temperatures shown. Additionally, the central value of the quantity increases only very slightly as the temperature increases. While quite small for these values of $T$, the effect has a physical explanation in the increase of the thermal population of the particles.

\begin{figure*}
    \centering
    \includegraphics[width=0.98\textwidth]{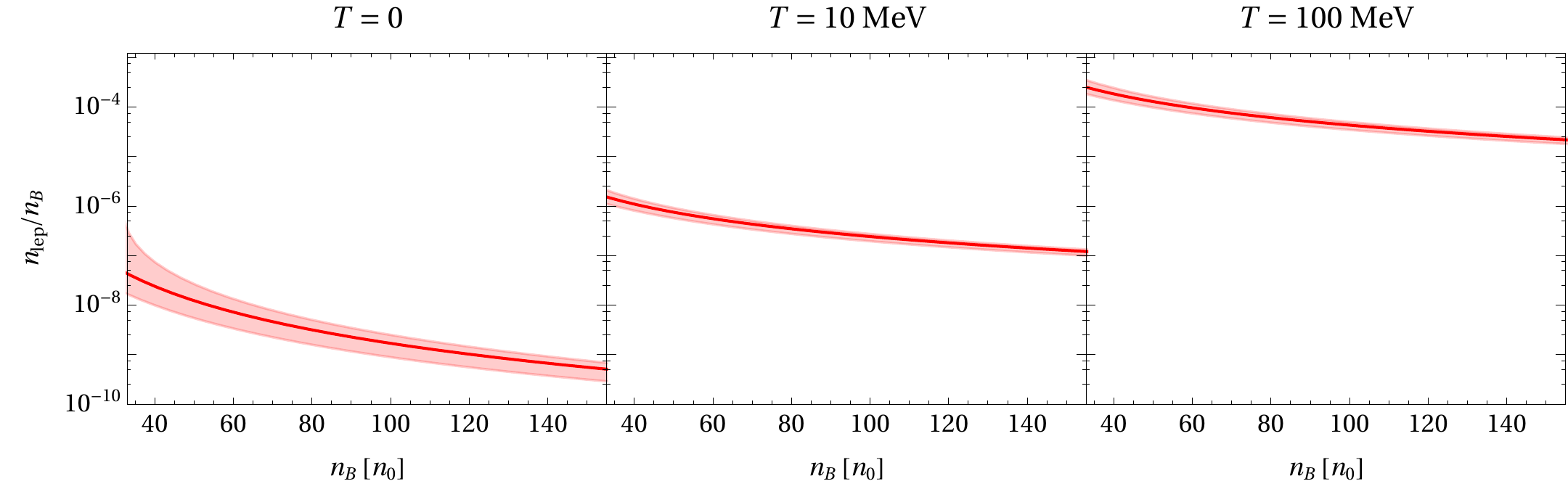}
    \caption{Left: A comparison of lepton fraction $n_\text{lep} / n_B$ at $T = 0$ (left panel), $T = 10$~MeV (middle panel), and $T = 100$~MeV (right panel)  for charge-neutral matter in beta equilibrium, ignoring neutrinos. Note that ignoring the strange quark mass would set $n_\text{lep} / n_{B} = 0$ under these conditions.}
    \label{fig:xlep_comparisons}
\end{figure*}
In \fig\ref{fig:mu_lep}, we show the progression of the lepton chemical potential as a function of baryon density under these conditions. As expected, the lepton chemical potential increases with decreasing  baryon density, as the quark chemical potentials decrease and the difference between $d$ and $s$ quarks become more pronounced. However, like the strangeness fraction above, $\mu_\text{lep}$ is only very modestly dependent on the temperature for $T \lesssim 200$~MeV at these densities, and for $T < 100$~MeV, relevant for NS mergers, thermal corrections to these quantities are negligible at these densities.  We note here that since the ratio $m_\mu / \mu_\text{lep}$ is not small, the muon mass must indeed be fully included within these computations, as has been done.

Finally, in \fig\ref{fig:xlep_comparisons}, we show the lepton fraction $n_\text{lep}/n_B$, with ${n_\text{lep} = n_e + n_\mu}$ as a function of baryon density under the same conditions. In this case, we see that the quantity deviates from its high-density limit of $0$ by less than $10^{-6}$ at all densities shown at $T = 0$, and as the temperature increases it grows quite strongly. This growth can be explained by the form of the number density of a massless lepton (for simplicity of explanation):
\begin{equation}
    \label{eq:massless_lepton_density}
    n_e(\mu_e, T) = \frac{\mu_e^3}{3 \pi ^2}+\frac{\mu_e T^2}{3}.
\end{equation}
The growth with temperature stems from the second term in \eq\eqref{eq:massless_lepton_density}, coupled with the fact that the lepton chemical potential is approximately temperature independent. We note that the crossover between $T = 0$ and $T$-dominated behavior for the lepton fraction occurs when $\pi T \approx \mu_\text{lep}$, which occurs for $T \approx 2$~MeV for the values of the $\mu_\text{lep}$ obtained here.

We turn now to a discussion of the results and methods presented in this work.
The methodology used above has some significant advantages over fully including the quark-mass effects. 

First, the relevant loop integrals are significantly simpler: Rather than containing the full mass dependence, they are zero-mass integrals with increasing powers of momenta in the denominator. This holds true to any finite-order expansion in the masses. With this in mind, considering masses larger in comparison to the density or temperature poses no fundamental problems so long as they still scale according to the coupling. For example, assuming the masses to be of the order $O(g^{1/2}\mu, g^{1/2} T)$ instead of the usual soft scale $O(g\mu, g T)$ considered here is a straightforward extension of our results. These improvements could for instance be carried out in the context of beyond-the-standard-model theories with additional heavy particles at high temperatures. Similarly, extension to higher orders in the coupling is relatively simple, as no fundamentally new integrals are required: expanded massive corrections to $n$-loop integrals are simply massless $n$-loop integrals with denominators raised to a higher power. For example, the 1-loop integrals $\widetilde{\mathcal{I}}^k_n$ in \eq\eqref{eq:Itilde} have a known closed-form expression for all $n$ and $k$, and in \app\ref{app:mass_exp_DR} we give results for contributions to $\mE^3$ in a form that immediately allows for evaluation to any order in the masses. Hence, as further perturbative corrections beyond $O(g^5)$ are calculated in the massless case, such as those following the program at $T=0$ recently advanced in \Refs\cite{Gorda:2021kme,Gorda:2021znl}, corrections from nonzero quark masses can then be included following our approach.

Second, our results themselves can be presented in a very simple form. Rather than involving complicated numerical integrals which are often badly behaved in certain regions of the parameter space, the results can be given in a closed analytic form, or in a simple numerical approximation. For this reason, our results have great potential for the community when it comes to applications to dense systems in astrophysics and other fields: The results incorporate the effects of a nonzero temperature and nonzero quark masses, allowing the study of phenomenologically interesting problems where these effects are necessary, yet they still maintain a simple and convenient form. For instance, with these results, one can extend the matching program of \Refs\cite{Annala:2017llu,Annala:2019puf} to constrain the NS equation of state to nonzero temperatures while including the effects of the strange quark mass.

As our numerical comparisons show, the advantages of our methodology come with only a minimal drop in accuracy compared to the full result, which we believe to be a more than acceptable tradeoff. 

\section*{Acknowledgements} We thank Aleksi Kurkela and Aleksi Vuorinen for helpful comments and suggestions.
TG was partially supported by the Deutsche Forschungsgemeinschaft (DFG, German Research Foundation) -- Project-Id 279384907 -- SFB 1245.

\appendix

\section{Unresummed contributions}
\label{app:unres_QCD}

As explained in the text, there are both matter and vacuum corrections from the unresummed QCD contributions, with the former being computable by expanding in the quark mass $m$. As each of the contributions computed in this Appendix only depends on a single chemical potential $\mu_f$ and mass $m_f$, we will drop the subscripts. In the main text, sums over flavors are displayed explicitly, and this will also be done in \app \ref{app:mass_exp_DR} to maintain clarity.

\subsection{Corrections from the free pressure}

The corrections to $\Omega_\text{free}$ from the quark mass are straightforward to compute by expanding 
\begin{align}
    \Omega_\text{free} =& -2 N_c \sumintF{P} \ln \left( P^2 + m^2 \right ) \\
    =& -2N_c\sumintF{P}  \left( \ln P^2 + \sum_{k=1}^{\infty} \frac{(-1)^{k+1}m^{2k}}{k(P^{2})^k}\right)
\end{align}
before integrating and are given in the main text. Here, we follow the notation of \Ref\cite{Vuorinen:2003fs}, with the fermionic sum-integral defined as
\begin{align}
    \sumintF{P} f(P^0,\vec{p}) &\equiv T \sum_{n = -\infty}^{\infty} \int_{\vec{p}} f\bigl((2n+1)\pi T - \text{i} \mu, \vec{p}\bigr), \\
    \int_{\vec{p}} &\equiv \left( \frac{e^{\gamma_\text{E}} \Lbar^2}{4 \pi} \right)^{\epsilon} \int \frac{{\rm d}^d \vec{p}}{(2 \pi)^d},
\end{align}
with $d = 3-2\epsilon$ and $\Lbar$ the $\overline{\text{MS}}$ renormalization scale.
The vacuum term simply comes from taking the unexpanded sum-integral above as a vacuum integral.

\subsection{Corrections from two-loop diagrams}

The correction to $\Omega_{2,\text{QCD}}$ requires dressing the one two-loop diagram containing quark lines with a mass correction. We find it easiest to first compute the scalarizations of this diagrams before expanding in the quark mass and then expand the scalar terms to the desired order in $m$.

The scalarization of the following diagram is a simple generalization of the scalarization provided in \Ref\cite{Kurkela:2009gj}\footnote{Note that \Ref\cite{Kurkela:2009gj} defines the diagrams as contributing to $-\Omega$}
\begin{align}
    \vcenter{\hbox{\includegraphics[height=1.3cm]{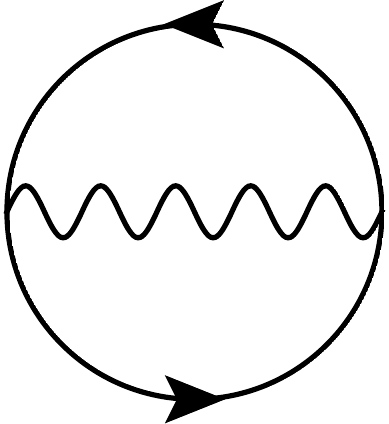}}} =& -g^2 \dA \left \{ \sumint_{\{\widetilde{P},\widetilde{Q}\}}   \frac{2 m^2}{(P^2 + m^2)(Q^2 + m^2)(P - Q)^2} \right. \nonumber \\
    & - (1 - \epsilon) \left [ \sumintF{P} \frac{1}{(P^2 + m^2)}\right]^2 \nonumber \\
    &\left. + 2(1-\epsilon) \sumint_{P} \frac{1}{P^2} \sumintF{Q} \frac{1}{(Q^2 + m^2)} \right\} \nonumber \\
    \equiv& I_1 + I_2 + I_3.
\label{eq:2loop_scalar}
\end{align}
Here, the bosonic sum-integral is defined as 
\begin{equation}
    \sumint_{P} f(P^0,\vec{p}) \equiv T \sum_{n = -\infty}^{\infty} \int_{\vec{p}} f\bigl(2n\pi T, \vec{p}\bigr).
\end{equation}
The pure matter parts are easy to compute in this form, where one must expand all terms to $O(m^2)$: 
\begin{align}
    I_{1} &= - 2 g^2 m^2 \dA \sumint_{\{\widetilde{P},\widetilde{Q}\}} \frac{1}{P^2 Q^2 (P-Q)^2} \displaybreak[0]\\
    I_{2} &= - 2 g^2 m^2 \dA (1-\epsilon) \sumint_{\{\widetilde{P},\widetilde{Q}\}} \frac{1}{P^2 (Q^2)^2 } \displaybreak[0]\\
    I_{3} &= 2 g^2 m^2 \dA (1-\epsilon) \sumint_{P,\{ \widetilde{Q}\}} \frac{1}{P^2 (Q^2)^2 }.
\end{align}
This reproduces \eq\eqref{eq:Domega_2}.

\section{Necessary integrals}
\label{app:ints}

We use the notation of \Ref\cite{Vuorinen:2003fs}  and define in the $\overline{\text{MS}}$ scheme in $d=3-2\epsilon$ dimensions
\begin{align}
    \label{eq:Ttilde}
    \widetilde{\mathcal{T}} \equiv& \sumint_{\{ \widetilde{P} , \widetilde{Q} \} } \frac{1}{P^2 Q^2 (Q - P)^2} \\
    =&-\frac{T^{2}}{\left(4\pi\right)^{2}}\Bigg\lbrace\left(\frac{\mu}{2\pi T}\right)^{2}\frac{1}{\epsilon}+\left(\frac{\mu}{2\pi T}\right)^{2}\left(2+4\ln\frac{\bar{\Lambda}}{4\pi T}\right) \nonumber \\
    &+\text{i}\frac{\mu}{\pi T}\lim_{s\rightarrow 0}\frac{\partial}{\partial s} \Bigg[\zeta\left(-s,\frac{1}{2}-\text{i}\frac{\mu}{2\pi T}\right) \nonumber \\
    &-\zeta\left(-s,\frac{1}{2}+\text{i}\frac{\mu}{2\pi T}\right)\Bigg] +O(\epsilon)\Bigg\rbrace \nonumber, \displaybreak[0]\\
    \mathcal{I}^k_n \equiv& \sumint_{Q} \frac{(Q^0)^k}{(Q^2)^{n}}\\
    =&\frac{\Gamma\left(n-\frac{3}{2}+\epsilon\right)\pi^{3/2}}{2\Gamma\left(n\right)}\left(\frac{e^{\gamma_{\mathrm{E}}}\bar{\Lambda}^{2}}{4\pi^{2}T^{2}}\right)^{\epsilon} \frac{T^{4-2n+k}}{\left(2\pi\right)^{2n-k}} \nonumber\\
    &\times \zeta\left(2n-k-3+2\epsilon\right),  \nonumber \displaybreak[0]\\
    \label{eq:Itilde}
    \widetilde{\mathcal{I}}^k_n \equiv& \sumint_{\{ \widetilde{Q} \}} \frac{(Q^0)^k}{(Q^2)^{n}} \\ =&\frac{\Gamma\left(n-\frac{3}{2}+\epsilon\right)\pi^{3/2}}{\Gamma\left(n\right)}\left(\frac{e^{\gamma_{\mathrm{E}}}\bar{\Lambda}^{2}}{4\pi^{2}T^{2}}\right)^{\epsilon} \frac{T^{4-2n+k}}{\left(2\pi\right)^{2n-k}} \nonumber\\
    &\times \Bigg[ \zeta\left(2n-k-3+2\epsilon,\frac{1}{2}-\text{i}\frac{\mu}{2\pi T}\right) \nonumber \\
    &+(-1)^k \zeta\left(2n-k-3+2\epsilon,\frac{1}{2}+\text{i}\frac{\mu}{2\pi T}\right)\Bigg]. \nonumber 
\end{align}

The correct massive generalization of \eq\eqref{eq:Itilde} is 
\begin{align}
    \widetilde{\mathcal{K}}^k_n \equiv& \sumint_{\{ \widetilde{Q} \}} \frac{(Q^0)^k}{(Q^2+m^2)^{n}},
\end{align}
which appears in the generalization of $\mE^2$ to nonzero quark mass. Unlike the zero-mass counterpart, this integral admits no closed-form representation in terms of standard special functions.  

\section{Small-mass expansion of the dimensionally reduced sector}
\label{app:mass_exp_DR}

Consider now the dimensionally reduced sector in more detail. The relevant expansion parameter is $\mE^2$, which is related to the one-loop gluon polarization tensor $\Pi^{00}$ via
\begin{equation} \label{eq:mEdef}
    \mE^2 = \lim_{p\rightarrow 0} \lim_{p_0\rightarrow 0}\Pi^{00}(p,p_0;m)+O(g^4).
\end{equation}
It contributes to the pressure through the $O(\mE^3)$-term, which reads 
\begin{equation} \label{eq:mEterm}
  \mE^3 \Omega_{\text{1,DR}} = -\frac{d_A}{12\pi} g^3 \alpha_{\text{E4}}^{3/2} - \frac{d_A}{2(4\pi)^3} g^5 \alpha_{\text{E4}}^{1/2} \alpha_{\text{E6}} + O(g^7,\epsilon).
\end{equation}
The $\alpha_{\text{E6}}$-term here will contribute massive corrections only at higher orders, and as such we only need the $\alpha_{\text{E4}}$-term at nonzero quark masses. However, for reasons related to regularization it is more convenient to, instead of working with the matching parameter $\alpha_{\text{E4}}$, work with $\mE^2$ directly for general $d$. This leads to the following definition for a nonzero-mass generalization of $\mE$ to first order in $g^2$, but for any $m_f$ and $d$: 
\begin{equation}\label{eq:mEmass}
\mE^2 \simeq (d-1) g^2 \left \{ (d-1) N_c \mathcal{I}_1^0 + \sum_f \left[ 2\widetilde{\mathcal{K}}_2^2 -\widetilde{\mathcal{K}}_1^0 \right] \right \},
\end{equation}
with the integrals $\mathcal{I}_n^m$ and $\widetilde{\mathcal{K}}_n^m$ defined in \app\ref{app:ints}. Here and below, we use a $``\simeq"$ to denote that the equations are valid only to the lowest order in $g^2$. That is to say, we use $\mE^2$ obtained by truncating the expansion in \eq\eqref{eq:mEdef} to only include the one-loop polarization tensor, which is all we need for our calculations. When numerically evaluating the contribution for arbitrary quark masses, it is convenient to first perform the frequency sums in $\widetilde{\mathcal{K}}_2^2 , \widetilde{\mathcal{K}}_0^1$. This results in the following (finite) one-dimensional integral expression for $\mE^2$ in $d=3$:
\begin{align}\label{eq:mEmassInt}
 \mE^2 |_{d=3} & \simeq \,\frac{g^2 T^2 }{2 \pi^2} \int_{0}^{\infty}\!\!\! \mathrm{d}x \Bigg\lbrace \frac{4N_c x}{e^x -1} + x^2 \sum_{f} e^{\sqrt{x^2+y_f^2}} \\ \nonumber
 \times&\Bigg[ \frac{e^{z_f}}{(1+e^{\sqrt{x^2+y_f^2}+z_f})^2}  
 + \frac{e^{-z_f}}{(1+e^{\sqrt{x^2+y_f^2}-z_f})^2} \Bigg] \Bigg\rbrace,
\end{align}
in which we have defined the dimensionless variables ${y_f\equiv m_f/T}$ and $z_f\equiv \mu_f/T$, and one observes the appearance of the distribution functions.  Note that this expression is consistent with the one given in \Ref\cite{Haque:2018eph} after taking the limit given in \eq\eqref{eq:mEdef} for the latter. 

We can see that $\mE^2$ is analytic in the squared quark masses, so that the expansion will lead to an $O(g^3m^2)$ term as expected. While it is possible to obtain at least the first necessary correction to $\Delta \mE^2 \equiv \mE^2 - \mEl^2$ by expanding \eq\eqref{eq:mEmassInt} by taking a derivative with respect to the quark masses and evaluating the resulting integrals analytically, a simpler method is simply to expand the integrals in \eq\eqref{eq:mEmass}, allowing us to simultaneously do everything for arbitrary $d$. In doing so, the zero-mass recurrence relation ${\widetilde{\mathcal{I}}_{n}^{m+2} = (2n-2-d) \widetilde{\mathcal{I}}_{n-1}^{m} /(2n-2)}$ is of great use. 

The result of this procedure can be expressed in a resummed form 
\begin{align}
    &\Delta \mE^2 \simeq g^2 T^2 \frac{2(d-1)\Gamma\left( 2-\frac{d}{2}\right)}{(d-2)\sqrt{\pi}}  \\ \nonumber 
     &\!\times \sum_f \mathrm{Re} \Bigg[ \zeta \bigg(2-d;\frac{1}{2} + \text{i}\frac{z_f}{2\pi}, \frac{y_f}{2\pi} \bigg)
    -\zeta \bigg(2-d,\frac{1}{2} + \text{i}\frac{z_f}{2\pi} \bigg) \Bigg] \\ \nonumber 
    &-g^2 T^2\frac{2\Gamma\left( 2-\frac{d}{2}\right)}{\sqrt{\pi}}    \sum_f\left(\frac{y_f}{2\pi}\right)^2  \mathrm{Re}\, \zeta \bigg(4-d;\frac{1}{2} + \text{i}\frac{z_f}{2\pi},\frac{y_f}{2\pi} \bigg),
\end{align}
where a generalization of the Hurwitz $\zeta$ function has been defined as (an analytic continuation of)
\begin{align}
    \zeta(s;z,a) &\equiv \sum_{n=0}^{\infty} \frac{1}{[(n+z)^2+a^2]^{s/2}} \\ \nonumber & = \sum_{n=0}^{\infty} \frac{2^n \Gamma \big(n+\frac{s}{2}\big)}{(2n)!!\Gamma \big(\frac{s}{2}\big)} (-a^2)^n \zeta(s+2n,z).
\end{align}
This series expansion allows one to directly compute the result to any finite order in the masses in a closed form for arbitrary $d$.

Whether one works with the four-dimensional sum-integrals or other approaches, the eventual result becomes very simple after neglecting the $O(\epsilon)$ corrections in $d=3-2\epsilon$ dimensions:
\begin{align}
   \Delta\mE^2 =& g^2 T^2 \frac{\Gamma\left( 2-\frac{d}{2}\right)}{2\sqrt{\pi}} (d-3)\sum_{f} \left(\frac{y_f}{2\pi}\right)^2 \\ \nonumber
    &\times\mathrm{Re}\,\zeta\Bigg(4-d;\frac{1}{2} + \text{i}\frac{z_f}{2\pi}\Bigg)+ O(g^4,g^2 m_f^4) \\ \nonumber
    =& - \frac{g^2}{4\pi^2} \sum_{f} m_f^2 + O(g^4,g^2 m_f^4,\epsilon). 
\end{align}
Upon substitution to \eq\eqref{eq:mEterm}, we immediately obtain \eq\eqref{eq:mEcorr}.  

\bibliography{Refs.bib}

\begin{thebibliography}{19}%
\makeatletter
\providecommand \@ifxundefined [1]{%
 \@ifx{#1\undefined}
}%
\providecommand \@ifnum [1]{%
 \ifnum #1\expandafter \@firstoftwo
 \else \expandafter \@secondoftwo
 \fi
}%
\providecommand \@ifx [1]{%
 \ifx #1\expandafter \@firstoftwo
 \else \expandafter \@secondoftwo
 \fi
}%
\providecommand \natexlab [1]{#1}%
\providecommand \enquote  [1]{``#1''}%
\providecommand \bibnamefont  [1]{#1}%
\providecommand \bibfnamefont [1]{#1}%
\providecommand \citenamefont [1]{#1}%
\providecommand \href@noop [0]{\@secondoftwo}%
\providecommand \href [0]{\begingroup \@sanitize@url \@href}%
\providecommand \@href[1]{\@@startlink{#1}\@@href}%
\providecommand \@@href[1]{\endgroup#1\@@endlink}%
\providecommand \@sanitize@url [0]{\catcode `\\12\catcode `\$12\catcode
  `\&12\catcode `\#12\catcode `\^12\catcode `\_12\catcode `\%12\relax}%
\providecommand \@@startlink[1]{}%
\providecommand \@@endlink[0]{}%
\providecommand \url  [0]{\begingroup\@sanitize@url \@url }%
\providecommand \@url [1]{\endgroup\@href {#1}{\urlprefix }}%
\providecommand \urlprefix  [0]{URL }%
\providecommand \Eprint [0]{\href }%
\providecommand \doibase [0]{https://doi.org/}%
\providecommand \selectlanguage [0]{\@gobble}%
\providecommand \bibinfo  [0]{\@secondoftwo}%
\providecommand \bibfield  [0]{\@secondoftwo}%
\providecommand \translation [1]{[#1]}%
\providecommand \BibitemOpen [0]{}%
\providecommand \bibitemStop [0]{}%
\providecommand \bibitemNoStop [0]{.\EOS\space}%
\providecommand \EOS [0]{\spacefactor3000\relax}%
\providecommand \BibitemShut  [1]{\csname bibitem#1\endcsname}%
\let\auto@bib@innerbib\@empty
\bibitem [{\citenamefont {Schwarz}(2003)}]{Schwarz:2003du}%
  \BibitemOpen
  \bibfield  {author} {\bibinfo {author} {\bibfnamefont {D.~J.}\ \bibnamefont
  {Schwarz}},\ }\bibfield  {title} {\bibinfo {title} {{The first second of the
  universe}},\ }\href {https://doi.org/10.1002/andp.200310010} {\bibfield
  {journal} {\bibinfo  {journal} {Annalen Phys.}\ }\textbf {\bibinfo {volume}
  {12}},\ \bibinfo {pages} {220} (\bibinfo {year} {2003})},\ \Eprint
  {https://arxiv.org/abs/astro-ph/0303574} {arXiv:astro-ph/0303574}
  \BibitemShut {NoStop}%
\bibitem [{\citenamefont {Fodor}\ and\ \citenamefont {Katz}(2010)}]{fodor}%
  \BibitemOpen
  \bibfield  {author} {\bibinfo {author} {\bibfnamefont {Z.}~\bibnamefont
  {Fodor}}\ and\ \bibinfo {author} {\bibfnamefont {S.~D.}\ \bibnamefont
  {Katz}},\ }\bibfield  {title} {\bibinfo {title} {{Lattice QCD and the Phase
  Diagram of Quantum Chromodynamics}},\ }in\ \href@noop {} {\emph {\bibinfo
  {booktitle} {Landolt-Börnstein Vol. 23: Relativistic Heavy Ion Physics}}},\
  \bibinfo {editor} {edited by\ \bibinfo {editor} {\bibfnamefont
  {R.}~\bibnamefont {Stock}}}\ (\bibinfo  {publisher} {Springer-Verlag Berlin
  Heidelberg},\ \bibinfo {year} {2010})\ \Eprint
  {https://arxiv.org/abs/0908.3341} {arXiv:0908.3341 [hep-ph]} \BibitemShut
  {NoStop}%
\bibitem [{\citenamefont {Busza}\ \emph {et~al.}(2018)\citenamefont {Busza},
  \citenamefont {Rajagopal},\ and\ \citenamefont {van~der
  Schee}}]{Busza:2018rrf}%
  \BibitemOpen
  \bibfield  {author} {\bibinfo {author} {\bibfnamefont {W.}~\bibnamefont
  {Busza}}, \bibinfo {author} {\bibfnamefont {K.}~\bibnamefont {Rajagopal}},\
  and\ \bibinfo {author} {\bibfnamefont {W.}~\bibnamefont {van~der Schee}},\
  }\bibfield  {title} {\bibinfo {title} {{Heavy Ion Collisions: The Big
  Picture, and the Big Questions}},\ }\href
  {https://doi.org/10.1146/annurev-nucl-101917-020852} {\bibfield  {journal}
  {\bibinfo  {journal} {Ann. Rev. Nucl. Part. Sci.}\ }\textbf {\bibinfo
  {volume} {68}},\ \bibinfo {pages} {339} (\bibinfo {year} {2018})},\ \Eprint
  {https://arxiv.org/abs/1802.04801} {arXiv:1802.04801 [hep-ph]} \BibitemShut
  {NoStop}%
\bibitem [{\citenamefont {Lattimer}\ and\ \citenamefont
  {Prakash}(2001)}]{Lattimer:2000nx}%
  \BibitemOpen
  \bibfield  {author} {\bibinfo {author} {\bibfnamefont {J.~M.}\ \bibnamefont
  {Lattimer}}\ and\ \bibinfo {author} {\bibfnamefont {M.}~\bibnamefont
  {Prakash}},\ }\bibfield  {title} {\bibinfo {title} {{Neutron star structure
  and the equation of state}},\ }\href {https://doi.org/10.1086/319702}
  {\bibfield  {journal} {\bibinfo  {journal} {Astrophys. J.}\ }\textbf
  {\bibinfo {volume} {550}},\ \bibinfo {pages} {426} (\bibinfo {year}
  {2001})},\ \Eprint {https://arxiv.org/abs/astro-ph/0002232}
  {arXiv:astro-ph/0002232} \BibitemShut {NoStop}%
\bibitem [{\citenamefont {Kajantie}\ \emph {et~al.}(2003)\citenamefont
  {Kajantie}, \citenamefont {Laine}, \citenamefont {Rummukainen},\ and\
  \citenamefont {Schroder}}]{Kajantie:2002wa}%
  \BibitemOpen
  \bibfield  {author} {\bibinfo {author} {\bibfnamefont {K.}~\bibnamefont
  {Kajantie}}, \bibinfo {author} {\bibfnamefont {M.}~\bibnamefont {Laine}},
  \bibinfo {author} {\bibfnamefont {K.}~\bibnamefont {Rummukainen}},\ and\
  \bibinfo {author} {\bibfnamefont {Y.}~\bibnamefont {Schroder}},\ }\bibfield
  {title} {\bibinfo {title} {{The Pressure of hot QCD up to g6 ln(1/g)}},\
  }\href {https://doi.org/10.1103/PhysRevD.67.105008} {\bibfield  {journal}
  {\bibinfo  {journal} {Phys. Rev. D}\ }\textbf {\bibinfo {volume} {67}},\
  \bibinfo {pages} {105008} (\bibinfo {year} {2003})},\ \Eprint
  {https://arxiv.org/abs/hep-ph/0211321} {arXiv:hep-ph/0211321} \BibitemShut
  {NoStop}%
\bibitem [{\citenamefont {Gorda}\ \emph
  {et~al.}(2021{\natexlab{a}})\citenamefont {Gorda}, \citenamefont {Kurkela},
  \citenamefont {Paatelainen}, \citenamefont {S\"appi},\ and\ \citenamefont
  {Vuorinen}}]{Gorda:2021znl}%
  \BibitemOpen
  \bibfield  {author} {\bibinfo {author} {\bibfnamefont {T.}~\bibnamefont
  {Gorda}}, \bibinfo {author} {\bibfnamefont {A.}~\bibnamefont {Kurkela}},
  \bibinfo {author} {\bibfnamefont {R.}~\bibnamefont {Paatelainen}}, \bibinfo
  {author} {\bibfnamefont {S.}~\bibnamefont {S\"appi}},\ and\ \bibinfo {author}
  {\bibfnamefont {A.}~\bibnamefont {Vuorinen}},\ }\bibfield  {title} {\bibinfo
  {title} {{Soft Interactions in Cold Quark Matter}},\ }\href
  {https://doi.org/10.1103/PhysRevLett.127.162003} {\bibfield  {journal}
  {\bibinfo  {journal} {Phys. Rev. Lett.}\ }\textbf {\bibinfo {volume} {127}},\
  \bibinfo {pages} {162003} (\bibinfo {year} {2021}{\natexlab{a}})},\ \Eprint
  {https://arxiv.org/abs/2103.05658} {arXiv:2103.05658 [hep-ph]} \BibitemShut
  {NoStop}%
\bibitem [{\citenamefont {Gorda}\ \emph
  {et~al.}(2021{\natexlab{b}})\citenamefont {Gorda}, \citenamefont {Kurkela},
  \citenamefont {Paatelainen}, \citenamefont {S\"appi},\ and\ \citenamefont
  {Vuorinen}}]{Gorda:2021kme}%
  \BibitemOpen
  \bibfield  {author} {\bibinfo {author} {\bibfnamefont {T.}~\bibnamefont
  {Gorda}}, \bibinfo {author} {\bibfnamefont {A.}~\bibnamefont {Kurkela}},
  \bibinfo {author} {\bibfnamefont {R.}~\bibnamefont {Paatelainen}}, \bibinfo
  {author} {\bibfnamefont {S.}~\bibnamefont {S\"appi}},\ and\ \bibinfo {author}
  {\bibfnamefont {A.}~\bibnamefont {Vuorinen}},\ }\bibfield  {title} {\bibinfo
  {title} {{Cold quark matter at N3LO: Soft contributions}},\ }\href
  {https://doi.org/10.1103/PhysRevD.104.074015} {\bibfield  {journal} {\bibinfo
   {journal} {Phys. Rev. D}\ }\textbf {\bibinfo {volume} {104}},\ \bibinfo
  {pages} {074015} (\bibinfo {year} {2021}{\natexlab{b}})},\ \Eprint
  {https://arxiv.org/abs/2103.07427} {arXiv:2103.07427 [hep-ph]} \BibitemShut
  {NoStop}%
\bibitem [{\citenamefont {Kurkela}\ and\ \citenamefont
  {Vuorinen}(2016)}]{Kurkela:2016was}%
  \BibitemOpen
  \bibfield  {author} {\bibinfo {author} {\bibfnamefont {A.}~\bibnamefont
  {Kurkela}}\ and\ \bibinfo {author} {\bibfnamefont {A.}~\bibnamefont
  {Vuorinen}},\ }\bibfield  {title} {\bibinfo {title} {{Cool quark matter}},\
  }\href {https://doi.org/10.1103/PhysRevLett.117.042501} {\bibfield  {journal}
  {\bibinfo  {journal} {Phys. Rev. Lett.}\ }\textbf {\bibinfo {volume} {117}},\
  \bibinfo {pages} {042501} (\bibinfo {year} {2016})},\ \Eprint
  {https://arxiv.org/abs/1603.00750} {arXiv:1603.00750 [hep-ph]} \BibitemShut
  {NoStop}%
\bibitem [{\citenamefont {Kurkela}\ \emph {et~al.}(2010)\citenamefont
  {Kurkela}, \citenamefont {Romatschke},\ and\ \citenamefont
  {Vuorinen}}]{Kurkela:2009gj}%
  \BibitemOpen
  \bibfield  {author} {\bibinfo {author} {\bibfnamefont {A.}~\bibnamefont
  {Kurkela}}, \bibinfo {author} {\bibfnamefont {P.}~\bibnamefont
  {Romatschke}},\ and\ \bibinfo {author} {\bibfnamefont {A.}~\bibnamefont
  {Vuorinen}},\ }\bibfield  {title} {\bibinfo {title} {{Cold Quark Matter}},\
  }\href {https://doi.org/10.1103/PhysRevD.81.105021} {\bibfield  {journal}
  {\bibinfo  {journal} {Phys. Rev. D}\ }\textbf {\bibinfo {volume} {81}},\
  \bibinfo {pages} {105021} (\bibinfo {year} {2010})},\ \Eprint
  {https://arxiv.org/abs/0912.1856} {arXiv:0912.1856 [hep-ph]} \BibitemShut
  {NoStop}%
\bibitem [{\citenamefont {Fraga}\ and\ \citenamefont
  {Romatschke}(2005)}]{Fraga:2004gz}%
  \BibitemOpen
  \bibfield  {author} {\bibinfo {author} {\bibfnamefont {E.~S.}\ \bibnamefont
  {Fraga}}\ and\ \bibinfo {author} {\bibfnamefont {P.}~\bibnamefont
  {Romatschke}},\ }\bibfield  {title} {\bibinfo {title} {{The Role of quark
  mass in cold and dense perturbative QCD}},\ }\href
  {https://doi.org/10.1103/PhysRevD.71.105014} {\bibfield  {journal} {\bibinfo
  {journal} {Phys. Rev. D}\ }\textbf {\bibinfo {volume} {71}},\ \bibinfo
  {pages} {105014} (\bibinfo {year} {2005})},\ \Eprint
  {https://arxiv.org/abs/hep-ph/0412298} {arXiv:hep-ph/0412298} \BibitemShut
  {NoStop}%
\bibitem [{\citenamefont {Laine}\ and\ \citenamefont
  {Schroder}(2006)}]{Laine:2006cp}%
  \BibitemOpen
  \bibfield  {author} {\bibinfo {author} {\bibfnamefont {M.}~\bibnamefont
  {Laine}}\ and\ \bibinfo {author} {\bibfnamefont {Y.}~\bibnamefont
  {Schroder}},\ }\bibfield  {title} {\bibinfo {title} {{Quark mass thresholds
  in QCD thermodynamics}},\ }\href {https://doi.org/10.1103/PhysRevD.73.085009}
  {\bibfield  {journal} {\bibinfo  {journal} {Phys. Rev. D}\ }\textbf {\bibinfo
  {volume} {73}},\ \bibinfo {pages} {085009} (\bibinfo {year} {2006})},\
  \Eprint {https://arxiv.org/abs/hep-ph/0603048} {arXiv:hep-ph/0603048}
  \BibitemShut {NoStop}%
\bibitem [{\citenamefont {Jim\'enez}\ and\ \citenamefont
  {Fraga}(2020)}]{Jimenez:2019kji}%
  \BibitemOpen
  \bibfield  {author} {\bibinfo {author} {\bibfnamefont {J.~C.}\ \bibnamefont
  {Jim\'enez}}\ and\ \bibinfo {author} {\bibfnamefont {E.~S.}\ \bibnamefont
  {Fraga}},\ }\bibfield  {title} {\bibinfo {title} {{Cold quark matter with
  heavy quarks and the stability of charm stars}},\ }\href
  {https://doi.org/10.1103/PhysRevD.102.034015} {\bibfield  {journal} {\bibinfo
   {journal} {Phys. Rev. D}\ }\textbf {\bibinfo {volume} {102}},\ \bibinfo
  {pages} {034015} (\bibinfo {year} {2020})},\ \Eprint
  {https://arxiv.org/abs/1908.10415} {arXiv:1908.10415 [hep-ph]} \BibitemShut
  {NoStop}%
\bibitem [{\citenamefont {Vuorinen}(2003)}]{Vuorinen:2003fs}%
  \BibitemOpen
  \bibfield  {author} {\bibinfo {author} {\bibfnamefont {A.}~\bibnamefont
  {Vuorinen}},\ }\bibfield  {title} {\bibinfo {title} {{The Pressure of QCD at
  finite temperatures and chemical potentials}},\ }\href
  {https://doi.org/10.1103/PhysRevD.68.054017} {\bibfield  {journal} {\bibinfo
  {journal} {Phys. Rev. D}\ }\textbf {\bibinfo {volume} {68}},\ \bibinfo
  {pages} {054017} (\bibinfo {year} {2003})},\ \Eprint
  {https://arxiv.org/abs/hep-ph/0305183} {arXiv:hep-ph/0305183} \BibitemShut
  {NoStop}%
\bibitem [{\citenamefont {Laine}\ and\ \citenamefont
  {Schroder}(2005)}]{Laine:2005ai}%
  \BibitemOpen
  \bibfield  {author} {\bibinfo {author} {\bibfnamefont {M.}~\bibnamefont
  {Laine}}\ and\ \bibinfo {author} {\bibfnamefont {Y.}~\bibnamefont
  {Schroder}},\ }\bibfield  {title} {\bibinfo {title} {{Two-loop QCD gauge
  coupling at high temperatures}},\ }\href
  {https://doi.org/10.1088/1126-6708/2005/03/067} {\bibfield  {journal}
  {\bibinfo  {journal} {JHEP}\ }\textbf {\bibinfo {volume} {03}},\ \bibinfo
  {pages} {067}},\ \Eprint {https://arxiv.org/abs/hep-ph/0503061}
  {arXiv:hep-ph/0503061} \BibitemShut {NoStop}%
\bibitem [{\citenamefont {Chakraborty}\ \emph {et~al.}(2015)\citenamefont
  {Chakraborty}, \citenamefont {Davies}, \citenamefont {Galloway},
  \citenamefont {Knecht}, \citenamefont {Koponen}, \citenamefont {Donald},
  \citenamefont {Dowdall}, \citenamefont {Lepage},\ and\ \citenamefont
  {McNeile}}]{Chakraborty:2014aca}%
  \BibitemOpen
  \bibfield  {author} {\bibinfo {author} {\bibfnamefont {B.}~\bibnamefont
  {Chakraborty}}, \bibinfo {author} {\bibfnamefont {C.~T.~H.}\ \bibnamefont
  {Davies}}, \bibinfo {author} {\bibfnamefont {B.}~\bibnamefont {Galloway}},
  \bibinfo {author} {\bibfnamefont {P.}~\bibnamefont {Knecht}}, \bibinfo
  {author} {\bibfnamefont {J.}~\bibnamefont {Koponen}}, \bibinfo {author}
  {\bibfnamefont {G.~C.}\ \bibnamefont {Donald}}, \bibinfo {author}
  {\bibfnamefont {R.~J.}\ \bibnamefont {Dowdall}}, \bibinfo {author}
  {\bibfnamefont {G.~P.}\ \bibnamefont {Lepage}},\ and\ \bibinfo {author}
  {\bibfnamefont {C.}~\bibnamefont {McNeile}},\ }\bibfield  {title} {\bibinfo
  {title} {{High-precision quark masses and QCD coupling from $n_f=4$ lattice
  QCD}},\ }\href {https://doi.org/10.1103/PhysRevD.91.054508} {\bibfield
  {journal} {\bibinfo  {journal} {Phys. Rev. D}\ }\textbf {\bibinfo {volume}
  {91}},\ \bibinfo {pages} {054508} (\bibinfo {year} {2015})},\ \Eprint
  {https://arxiv.org/abs/1408.4169} {arXiv:1408.4169 [hep-lat]} \BibitemShut
  {NoStop}%
\bibitem [{\citenamefont {Alford}\ and\ \citenamefont
  {Harris}(2018)}]{Alford:2018lhf}%
  \BibitemOpen
  \bibfield  {author} {\bibinfo {author} {\bibfnamefont {M.~G.}\ \bibnamefont
  {Alford}}\ and\ \bibinfo {author} {\bibfnamefont {S.~P.}\ \bibnamefont
  {Harris}},\ }\bibfield  {title} {\bibinfo {title} {{Beta equilibrium in
  neutron star mergers}},\ }\href {https://doi.org/10.1103/PhysRevC.98.065806}
  {\bibfield  {journal} {\bibinfo  {journal} {Phys. Rev. C}\ }\textbf {\bibinfo
  {volume} {98}},\ \bibinfo {pages} {065806} (\bibinfo {year} {2018})},\
  \Eprint {https://arxiv.org/abs/1803.00662} {arXiv:1803.00662 [nucl-th]}
  \BibitemShut {NoStop}%
\bibitem [{\citenamefont {Annala}\ \emph {et~al.}(2018)\citenamefont {Annala},
  \citenamefont {Gorda}, \citenamefont {Kurkela},\ and\ \citenamefont
  {Vuorinen}}]{Annala:2017llu}%
  \BibitemOpen
  \bibfield  {author} {\bibinfo {author} {\bibfnamefont {E.}~\bibnamefont
  {Annala}}, \bibinfo {author} {\bibfnamefont {T.}~\bibnamefont {Gorda}},
  \bibinfo {author} {\bibfnamefont {A.}~\bibnamefont {Kurkela}},\ and\ \bibinfo
  {author} {\bibfnamefont {A.}~\bibnamefont {Vuorinen}},\ }\bibfield  {title}
  {\bibinfo {title} {{Gravitational-wave constraints on the neutron-star-matter
  Equation of State}},\ }\href {https://doi.org/10.1103/PhysRevLett.120.172703}
  {\bibfield  {journal} {\bibinfo  {journal} {Phys. Rev. Lett.}\ }\textbf
  {\bibinfo {volume} {120}},\ \bibinfo {pages} {172703} (\bibinfo {year}
  {2018})},\ \Eprint {https://arxiv.org/abs/1711.02644} {arXiv:1711.02644
  [astro-ph.HE]} \BibitemShut {NoStop}%
\bibitem [{\citenamefont {Annala}\ \emph {et~al.}(2020)\citenamefont {Annala},
  \citenamefont {Gorda}, \citenamefont {Kurkela}, \citenamefont {N\"attil\"a},\
  and\ \citenamefont {Vuorinen}}]{Annala:2019puf}%
  \BibitemOpen
  \bibfield  {author} {\bibinfo {author} {\bibfnamefont {E.}~\bibnamefont
  {Annala}}, \bibinfo {author} {\bibfnamefont {T.}~\bibnamefont {Gorda}},
  \bibinfo {author} {\bibfnamefont {A.}~\bibnamefont {Kurkela}}, \bibinfo
  {author} {\bibfnamefont {J.}~\bibnamefont {N\"attil\"a}},\ and\ \bibinfo
  {author} {\bibfnamefont {A.}~\bibnamefont {Vuorinen}},\ }\bibfield  {title}
  {\bibinfo {title} {{Evidence for quark-matter cores in massive neutron
  stars}},\ }\href {https://doi.org/10.1038/s41567-020-0914-9} {\bibfield
  {journal} {\bibinfo  {journal} {Nature Phys.}\ }\textbf {\bibinfo {volume}
  {16}},\ \bibinfo {pages} {907} (\bibinfo {year} {2020})},\ \Eprint
  {https://arxiv.org/abs/1903.09121} {arXiv:1903.09121 [astro-ph.HE]}
  \BibitemShut {NoStop}%
\bibitem [{\citenamefont {Haque}(2018)}]{Haque:2018eph}%
  \BibitemOpen
  \bibfield  {author} {\bibinfo {author} {\bibfnamefont {N.}~\bibnamefont
  {Haque}},\ }\bibfield  {title} {\bibinfo {title} {{Quark mass dependent
  collective excitations and quark number susceptibilities within the hard
  thermal loop approximation}},\ }\href
  {https://doi.org/10.1103/PhysRevD.98.014013} {\bibfield  {journal} {\bibinfo
  {journal} {Phys. Rev. D}\ }\textbf {\bibinfo {volume} {98}},\ \bibinfo
  {pages} {014013} (\bibinfo {year} {2018})},\ \Eprint
  {https://arxiv.org/abs/1804.04996} {arXiv:1804.04996 [hep-ph]} \BibitemShut
  {NoStop}%
\end{thebibliography}%

\end{document}